# Nonlinear Spectroscopy and All-Optical Switching of Femtosecond Soliton Molecules

F. Kurtz[1], C. Ropers[1], G. Herink[2*]

[1] IV. Physical Institute - Solids and Nanostructures, University of Göttingen, Germany
[2] Experimental Physics VIII - Ultrafast Dynamics, University of Bayreuth, Bayreuth, Germany

**Abstract:** The emergence of confined structures and pattern formation are exceptional manifestations of concurring nonlinear interactions found in a variety of physical, chemical and biological systems[1]. Optical solitons are a hallmark of extreme spatial or temporal confinement enabled by a variety of nonlinearities. Such particle-like structures can assemble in complex stable arrangements, forming "soliton molecules"[2,3]. Recent works revealed oscillatory internal motions of these bound states, akin to molecular vibrations[4–8]. These observations beg the question as to how far the "molecular" analogy reaches, whether further concepts from molecular spectroscopy apply in this scenario, and if such intra-molecular dynamics can be externally driven or manipulated. Here, we probe and control such ultrashort bound-states in an optical oscillator, utilizing real-time spectroscopy and time-dependent external perturbations. We introduce two-dimensional spectroscopy of the linear and nonlinear bound-state response and resolve anharmonicities in the soliton interaction leading to overtone and sub-harmonic generation. Employing a non-perturbative interaction, we demonstrate all-optical switching between distinct states with different binding separation, opening up novel schemes of ultrafast spectroscopy, optical logic operations and all-optical memory.

Mirroring the hierarchy of the atomic and molecular composition of matter, fundamental solitons constitute stable entities which can aggregate to form structures of increased complexity. Such condensation manifests in soliton fluids, molecules and crystals, as observed in diverse physical systems[2,9,10]. The underlying forces are responsible for various emergent phenomena, such as self-organization of Bose-Einstein condensates, soliton crystallization in micro-cavities, soliton fission in supercontinua or rogue waves in optical fiber[10–15]. It is well-known that ultrashort solitons can form highly-stable bound-states, and these optical soliton molecules arise from balanced forces during propagation in passive fibers and in dissipative laser resonators[2–5,8,16–21]. Recent real-time studies have revealed rapid transients, internal vibrations and even chaotic dynamics[6,7,22–24].

In this study, we transfer the concepts of optical spectroscopy to the case of ultrafast soliton molecules by driving them with an external perturbation and monitoring their response in real time. We discover a resonance in the response of soliton molecules to external perturbation and probe the associated anharmonic binding potential (Fig. 1a, approach I). Applying stronger, non-perturbative stimuli, we also demonstrate reliable and reversible all-optical switching between two different bound states (Fig. 1a, approach II) – suggesting applications in fast optical sampling and pulse control.

Correspondence: *georg.herink@uni-bayreuth.de

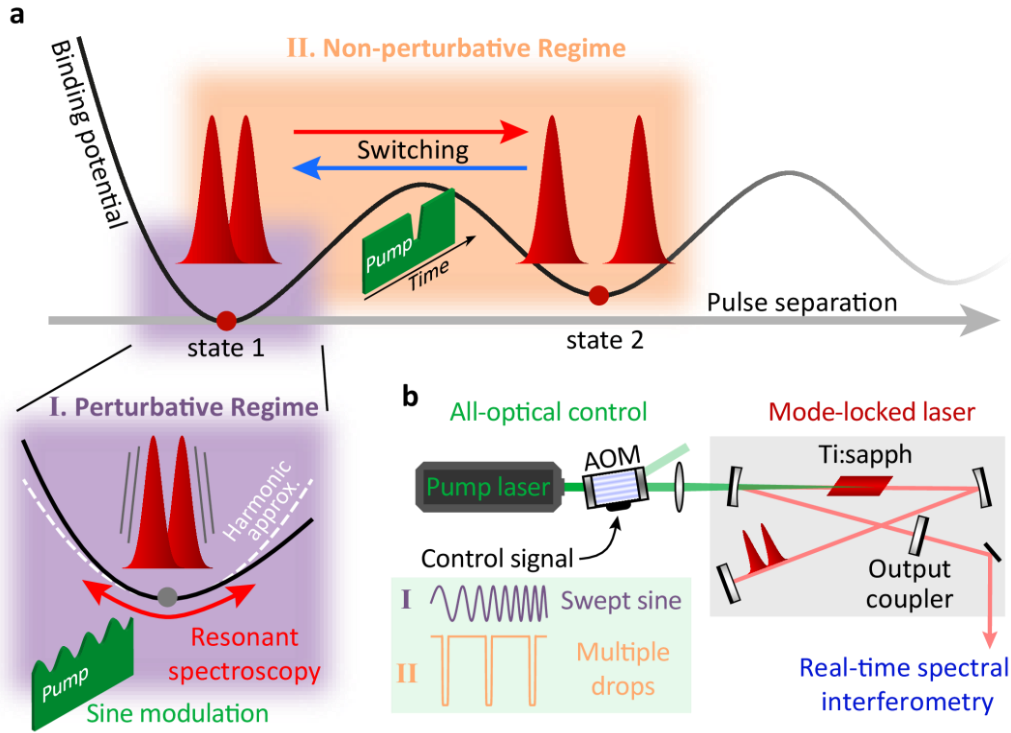

**Figure 1 | All-optical Control of Soliton Molecules a)** Soliton molecules can exhibit different stable bound states, as represented by a set of local minima in their binding potential. These states are perturbed via modulating the pump power in two ways: *(I.)* Sinusoidal modulation drives the bound states into resonant vibration. The resonant response and anharmonic contributions are probed by frequency-swept excitation. *(II.)* The application of a rapid transient drop in pump power allows for all-optical switching between two bound states of different pulse separation. **b)** All-optical control is experimentally implemented using an acousto-optic modulator (AOM) in the pump beam of the laser oscillator. Control patterns are generated from arbitrary electronic waveforms driving the AOM.

We conduct experiments on a commercial Kerr-lens mode-locked Ti:sapphire oscillator supporting ultrashort pulses of sub-10 fs duration. External all-optical control of the soliton dynamics is facilitated via rapid modulations of the pump power by acousto-optic modulation (AOM) of the pump beam, as sketched in Fig. 1b. The bandwidth of the AOM exceeds 1 MHz, and its response was carefully characterized in order to linearize the transfer function between the control input and optical output signals. After every round trip in the laser cavity, the out-coupled soliton molecules are measured in real-time by employing spectral interference and single-shot detection at a framerate of 78 MHz, utilizing the time-stretch dispersive Fourier transform (TS-DFT)[6,25]. This scheme reveals shot-to-shot changes in the temporal separation $\tau$ and the relative phase $\Delta\varphi$ of the solitons with envelopes $E_1(t), E_2(t)$ within the molecule, $E(t) = Re\{(E_1(t) + E_2(t))\exp(i\omega_0 t)\}$ with $E_2(t) = E_1(t + \tau)\exp(i\Delta\varphi)$ (see sketch in Fig. 2a and Ref. [6]). Synchronized with the pump modulation, this allows us to conduct "soliton molecule spectroscopy", i.e., a frequency-dependent characterization of the bound-state response to external perturbation.

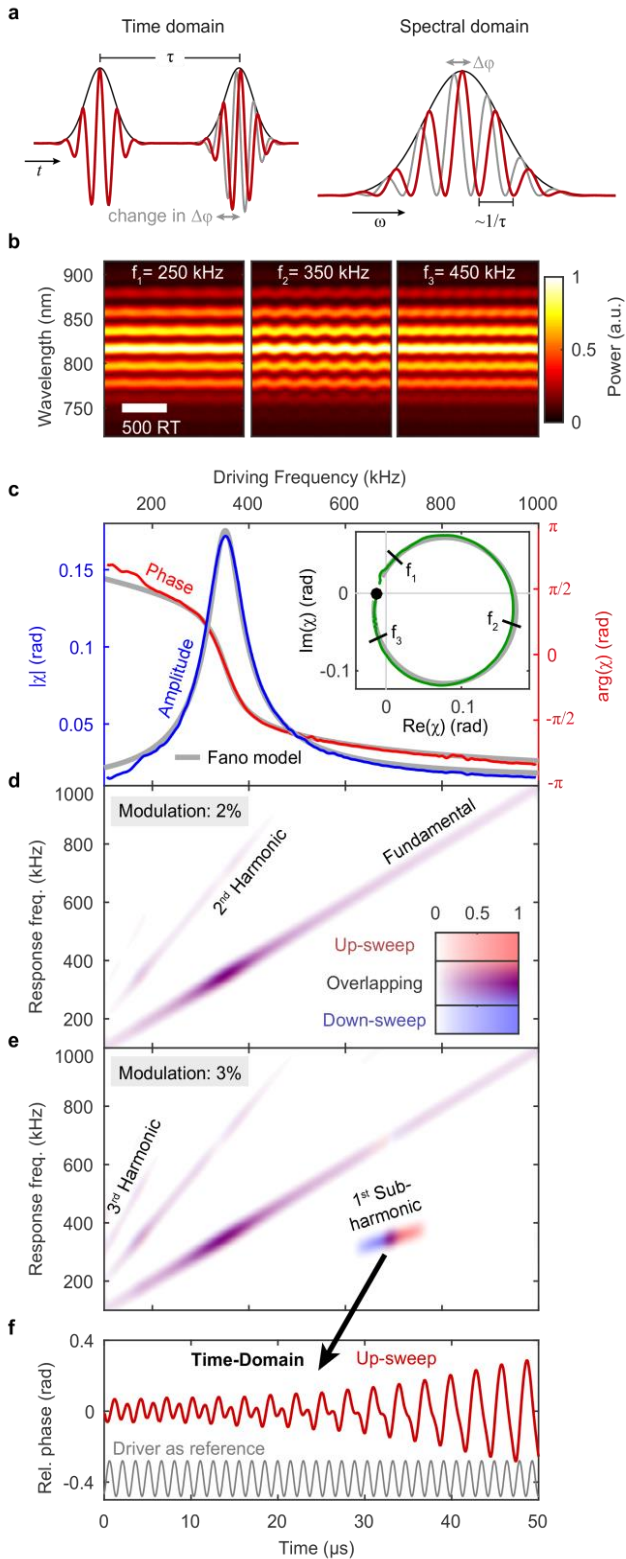

**Figure 2 | Driven Soliton Molecules in the Perturbative Regime a)** The temporal separation and relative phase of the solitons within the molecule are visible by the pitch and location, respectively, of the fringe period in the frequency domain. **b)** Three segments of experimental real-time spectra during frequency-swept excitation, displaying resonant behavior at $f_2$=350 kHz. Scale bar in roundtrips (RT). **c)** The linear response (amplitude and phase) extracted from the frequency sweep. The susceptibility $\chi(f)$ is described by a Fano resonance (grey lines). Inset: Corresponding Nyquist plot in the complex plane. **d)** 2D-spectra extracted from short-time Fourier transforms of up- and down-sweeps. Colormap indicates intensity for up-sweep (red), down-sweep (blue) and overlapping sweeps (purple). At 2% modulation, both traces overlap, and additional harmonics are excited. **e)** Stronger modulation (3%) enhances the harmonics and drives sub-harmonic motion with pronounced hysteresis (see non-overlapping up- and down-sweeps). **f)** The real-time waveform displays the transition into sub-harmonic oscillation during the up-sweep (red, driving signal as reference in grey).

**Spectroscopy in the perturbative regime**

In a first experiment, we study soliton molecules under weak and intermediate excitation. To this end, we prepare a stationary soliton molecule with locked relative phase as the initial state. The dynamic response is probed for harmonic modulation of the pump power $P(t) = P_0(1 + M_f \cos(2\pi f t))$, where $P_0$ is the average power, $M_f$ is the relative modulation, and the frequency $f$ is adiabatically swept from *100* kHz to *1* MHz. Representative segments of real-time spectral interferograms are presented in Fig. 2b. The soliton molecule primarily responds by oscillations in the relative phase $\Delta\varphi(t)$ around a value $\Delta\varphi_0$, as evident from the periodically shifting fringes. We find that the frequency-dependent amplitude of these oscillations exhibits a resonance behavior peaked at $f_2 = 350$ kHz. Evaluating the phase oscillation across the entire frequency sweep, we extract the complex linear susceptibility $\chi(f) = (\Delta\varphi(f) - \Delta\varphi_0)/M_f$, which quantifies the frequency-dependent oscillations of $\Delta\varphi$ (in amplitude and 'phase') relative to the modulation strength (see Fig. 2c). We find that the susceptibility is well described by a Fano-type lineshape[26] (grey lines)

$$\chi(f) = \tilde{a}_{nr} + \tilde{a}_r \frac{\Gamma}{f - f_0 + i\Gamma},$$

with an asymmetric profile $|\chi(f)|$ due to the interference of resonant ($\tilde{a}_r = (0.042 + 0.183i)$ rad) and non-resonant ($\tilde{a}_{nr} = -(0.0135 + 0.0012i)$ rad) contributions. Here, $f_0 = 350$ kHz is the resonance frequency and $\Gamma = 32$ kHz the damping constant. Generally, changes in the pump power couple to the soliton energies, and this perturbation induces a variation in the relative phase via the Kerr nonlinearity. Due to its instantaneous response, this subtle effect manifests as a spectrally flat, non-resonant contribution. The dominant resonant behavior can be understood by the proximity of the system to a supercritical Hopf bifurcation[27]. Such a bifurcation marks a transition between stationary and oscillatory states, controlled by a characteristic damping parameter. In the case of dissipative soliton molecules, the cw-pump power $P_0$ can serve as this control parameter: If $P_0$ drops below a critical value, spontaneous and self-sustained internal oscillations emerge, as previously observed[4–7,27]. Above the critical damping, typically only stationary soliton molecules are observed. However, as shown here, the underlying damped resonance is still amenable to spectroscopy.

Further insight into the bound-state response is provided by windowed Fourier transforms of $\Delta\varphi(t)$ during the sweep. Specifically, the set of Fourier spectra along the frequency-sweep forms a two-dimensional map (Fig. 2d), representing the complete response for a given excitation frequency. In addition to the dominant linear response, we observe distinct signatures of nonlinear dynamics, i.e., the excitation of 2nd and 3rd harmonic frequencies. Via increasing the modulation strength (Fig. 2e), the amplitudes of higher harmonics increase and, remarkably, a subharmonic ($f/2$)-oscillation can be triggered by excitation near $2f_0$. This ($f/2$)-oscillation is a result of a strong second-order nonlinearity in the restoring force. It implies the existence of two generic amplitude thresholds in the system[28]: Above the *necessary* driving threshold, damping of the subharmonic motion is surpassed. At the *sufficient* threshold, the fundamental oscillation ultimately destabilizes and subharmonic motion dominates. Such progressive conditions generate a hysteresis in the frequency-dependent amplitude response, as we indeed find in the non-overlapping bi-directional sweeps displayed in Fig. 2e (up-/down-sweeps in red/blue, overlapping in purple). Around the subharmonic resonance, the system preserves the information on the sweep direction, effectively representing optical memory of the trajectory. A time-domain view on the subharmonic buildup during the sweep is shown in Fig. 2f.

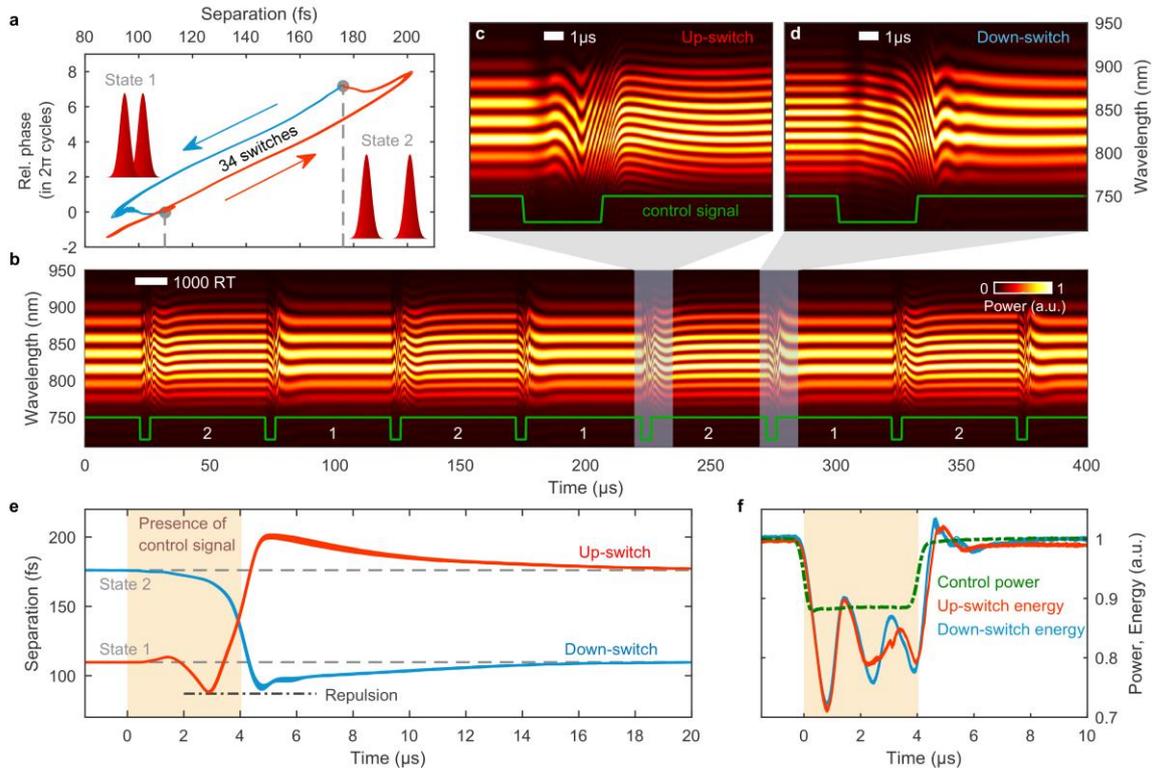

**Figure 3 | All-optical switching of soliton molecules in the non-perturbative regime: a)** Multiple trajectories of consecutive switching events between state 1 (small separation) and state 2 (large separation) presented in the separation-phase plane, demonstrating high switching fidelity. **b)** Corresponding real-time spectra with applied control signal sketched in green (arb.u.). Rapid spectral modulations indicate large soliton separation. **c,d)** Extended views on exemplary up- and down-switching, respectively. **e)** Evolution of the soliton separation during the switching events (17 events each). **f)** Rapid pump power drop (green) mediates characteristic transients in pulse energy (red/blue curves for up/down switching), which determines the switching behavior.

## All-optical switching

Beyond the perturbative probing of linear and nonlinear responses, we investigate the reaction of soliton molecules to more invasive transient changes in pump power. Specifically, we have identified that short-term notches in $P_0$ can be used to controllably switch the binding distance. Figures 3b-d display experimental real-time interferograms along with the control signal used. A rapid drop of 10% in the pump power disrupts the bound-state from its binding separation of $\tau_1 = 110$ fs (state 1). Initially, the molecule destabilizes, both pulses repel and separate (red curves, Fig. 3e). Upon recovery of the pump power, a phase-stable bound-state is gradually reestablished – yet at larger separation of $\tau_2 = 175$ fs (state 2). Notably, the subsequent application of an identical further stimulus triggers the transition back to the initial state 1, and, thus, facilitates a reversal to the former bound-state separation $\tau_1$. As a result, the application of multiple control stimuli enables a reversible and deterministic switching between two phase-stable doublet states. The individual switching processes evolve in a highly reproducible manner, as evidenced by multiple consecutive cycles tracing out the same trajectories in the configuration space spanned by $\tau$ and $\Delta\varphi$ (Fig. 3a). Both up- and down-switching involve a rapid transient collapse of the initial state separation, followed by a slower relaxation into the final state. Whereas the partial bound-state collapse is initiated by the global change of intracavity energy, its further evolution is conversely influenced by the internal interactions of the pulse pair. This feedback results in the deviating evolution of the intracavity energy for both switching directions, particularly evident 2 µs after the pump drop (cf. Fig. 3f).

The practically error-free switching is verified by observing the fidelity of the final state after application of several thousand events in a separate measurement. Our results demonstrate the realization of all-optical pulse control in the oscillator via dissipative soliton dynamics. Approaching MHz switching speeds, this scheme presents a new route to all-optical memory, relating to previous ultrafast laser implementations of multi-soliton manipulation[29–31].

In conclusion, we demonstrate two distinct regimes of external, all-optical control of soliton-soliton interactions. Our study extends the analogy of matter-like soliton molecules by translating general concepts of optical spectroscopy to the realm of temporal dissipative solitons. For moderate excitation, we probe the natural vibration resonance and reveal anharmonicities in the underlying binding potential giving rise to harmonic and subharmonic oscillations in the relative phase. The observed linewidth serves as a direct measure for the intrinsic vibrational damping, and tracking it as a function of pump power across the Hopf-bifurcation will be an interesting subject of further study. Applying a stronger, non-perturbative stimulus, we are able to deterministically induce transitions between distinct bound states. Initial observations also indicate access to larger sets of discrete binding states, highlighting the prospects of rapid multi-state switching of pulse pairs for ultrafast spectroscopy and for logic operations including counting, memory and all-optical soliton shift registers. Generally, we believe that soliton molecule spectroscopy will prove as a versatile tool to study and harness multi-soliton interactions in a broad range of ultrafast nonlinear systems.

**Author contributions**

All authors contributed equally to this work.